# CrossRoI: Cross-camera Region of Interest Optimization for Efficient Real Time Video Analytics at Scale

Hongpeng Guo
hg5@illinois.edu
UIUC

Shuochao Yao
shuochao@gmu.edu
George Mason University

Zhe Yang
zheyang3@illinois.edu
UIUC

Qian Zhou
qianz@illinois.edu
UIUC

Klara Nahrstedt
klara@illinois.edu
UIUC

## ABSTRACT

Video cameras are pervasively deployed in city scale for public good or community safety (i.e. traffic monitoring or suspected person tracking). However, analyzing large scale video feeds in real time is data intensive and poses severe challenges to today's network and computation systems. We present CrossRoI, a resource-efficient system that enables real time video analytics at scale via harnessing the videos content associations and redundancy across a fleet of cameras. CrossRoI exploits the intrinsic physical correlations of cross-camera viewing fields to drastically reduce the communication and computation costs. CrossRoI removes the repentant appearances of same objects in multiple cameras without harming comprehensive coverage of the scene. CrossRoI operates in two phases - an offline phase to establish cross-camera correlations, and an efficient online phase for real time video inference. Experiments on real-world video feeds show that CrossRoI achieves 42% ~ 65% reduction for network overhead and 25% ~ 34% reduction for response delay in real time video analytics applications with more than 99% query accuracy, when compared to baseline methods. If integrated with SotA frame filtering systems, the performance gains of CrossRoI reach 50% ~ 80% (network overhead) and 33% ~ 61% (end-to-end delay).

## 1 INTRODUCTION

Driven by plummeting camera prices and advances of intelligent video inference algorithms, video cameras are being deployed ubiquitously in recent days. For example, many cities in the world now deploy tens of thousands of cameras at key locations, such as highway entrances or roads intersections, to collect rich video data for applications ranging from traffic monitoring, public safety and suspected target tracking [1, 2, 4, 5, 42]. As tremendous data are being generated by the cameras in every second, organizations usually rely on live video analytics to retrieve key information, such as objects locations and identities, in real time. Two key enablers for fast and accurate video inference are the rapid development of deep neural networks, especially Convolutional Neural Networks (CNNs), and their hardware accelerators which empower fast and large-scale neural networks training and inference.

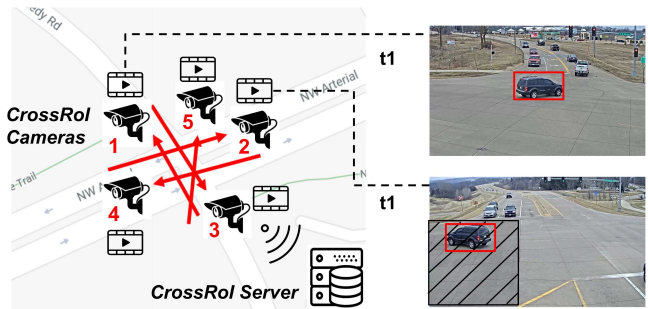

Figure 1: An application scenario of CrossRoI. Red arrows show the viewing angles of cameras. The two frames on the right are captured at timestamp $t_1$ from camera 1 and 2, respectively.

However, live video analytics in large scale are usually *network-exhaustive* and *compute-intensive*. In a typical video analytics pipeline, real time video feeds from widely deployed cameras are streamed to a cloud server or geographically close edge clusters where powerful hardware (e.g. GPUs) and fine-trained CNNs (e.g. YOLO [34]) are prepared. The server immediately loads videos into the inference pipelines and aims for accurate and low latency analytic results. The high network demands for video streaming and computation demands for CNN-based inference pose severe challenges to such video analytics framework, especially when organizations are steadily increasing their deployment scale, which amplifies the problems.

Significant work has been presented to improve the efficiency of video analytics pipelines, which can be categorized into two groups: (1) frame filtering on single camera [13, 14, 23, 27], and (2) target oriented cross-camera analytics [24, 25, 30]. Works in group one (e.g. Reducto [27]) optimize the cost/accuracy tradeoffs of single-video analytics with frame sampling or CNN-based filters for discarding frames. Reductos' optimizations are within single video streams independent of other streams, resulting in linear growth of resource demands and limitations to scaling. Works in group two (e.g. Spatula [25]) schedule the on and off of geographically distributed cameras to track a predefined target object across cameras. While Spatula substantially reduces network/computation demands by turning off the majority of cameras at any time, it fails to provide a comprehensive coverage to every scene where cameras are deployed.

In this work, we present CrossRoI **system** to address the resource-intensive challenges for real time video analytics on a fleet of closely located cameras (e.g. the cameras installed at a traffic intersection) via harnessing the video content associations and redundancy across the group of cameras. As shown in figure 1, 5 cameras are deployed at a road intersection with their viewing field overlapped. An object in the scene may appear in the *field-of-view* of multiple cameras at the same time. In many video analytics tasks (e.g. vehicle or suspect person detection), any capture of the interesting target is effective to fulfill the mission. For example in figure 1 either detection of the black car in camera 1 or 2 is enough to locate it at this traffic crossing at this moment ($t_1$). Removing the lower left region (shadow region) of camera 2's frame at $t_1$ does not influence the comprehensiveness of vehicle detection results at all [1]. We argue that both network traffic and computation demands can be substantially reduced without harming inference accuracy for video analytics pipelines if the intrinsic associations across cameras could be discovered and harnessed properly.

CrossRoI highlights three **challenges** to discover the intrinsic data associations and harness the content redundancy across multiple cameras as follows.

(**C1**) How to establish data association among a fleet of cameras on unlabeled video data *automatically* and *accurately*?
(**C2**) How to calculate cross-camera region-of-interest (RoI) collectively to *remove redundancy* without harming the *comprehensiveness* of detection coverage?
(**C3**) How to leverage cameras' regions-of-interest to drastically *reduce network overhead* and *boost sever inference throughput* in the video analytics pipeline?

To tackle these challenges, we design CrossRoI to operate in two distinct phases-an online phase and an offline phase. In the offline phase, CrossRoI establishes the data association and calculates the optimized RoI information. In the online phase, cameras filter their real time video streams according to the RoI information to reduce overall system data intensity. To establish cross-camera data association, we augment existing re-identification solutions with statistical filters to generate highly-accurate ReID results, and hence, use the cross-camera appearances of same objects to represent data correlations among cameras (**C1**). We slice camera frames into fine-grained tiles and develop a combinatorial optimization framework to calculate least-sized regions of interest among the camera fleet collectively without missing detection of any object (**C2**). To best alleviate resource-intensive challenges, we apply the optimized RoI information in each camera as a filter to prevent non-interesting data being dumped into the analytics pipeline. We further specially design video compression module and RoI based CNN inference pipeline to boost overall system performance(**C3**). Overall, this paper makes the following **contributions**:

(1) We augment existing re-identification (ReID) algorithms to establish cross-camera data association automatically and accurately.
(2) We develop an optimization framework to harness cross-camera data redundancy and significantly reduce the data intensity of the video analytics pipeline.
(3) Our specially designed video compression module and RoI-based CNN inference pipeline boost the overall system performance even further.
(4) Evaluations on real-world traffic videos suggest our system achieves network overhead reduction up to 65% and end-to-end response latency reduction by 34% compared to baselines.
(5) Compared to most frame-filtering based existing solutions (e.g., Reducto [27]), CrossRoI exemplifies an extra layer of optimization from spatial domain. When integrated with frame-filtering module, CrossRoI outperforms original frame-filtering systems by 2×, and outperforms baselines up to 5×, in terms of network usage.

The rest of the paper is organized as follows. In §2, we survey related literature and present backgrounds about video analytics, streaming and ReID frameworks. In §3, we present our ReID based cross-camera data associations and optimization framework to generate RoI masks for each camera. We present CrossRoI system workflow and design details in §4. §5 shows our evaluations of CrossRoI system. Finally, §6 concludes the paper.

## 2 BACKGROUNDS AND RELATED WORKS
### 2.1 Video Analytics Systems

Video analytics systems have been widely studied in recent literature. [11, 13–16, 23–25, 27, 28, 30, 33, 41, 45–47]. While all these works focus on solving the resource-intensive challenge, different approaches have been proposed. We categorize all existing works in terms of (1) their system architectures, (2) capabilities to fulfill real-time processing, and (3) processing multiple-camera videos independently or collectively, as follows.

Most works fells in either a three-layered *camera-edge-cloud* [11, 14, 23, 25, 28, 41, 45, 47] or a two-layered *camera-cloud* [13, 15, 20, 27, 33] architecture. The first class of work, exemplified by Focus [23], deploys close-to-camera edge devices to augment the processing power of cameras, and hence, prune redundant data using neural networks accurately before sending videos to the cloud for deep analysis. While the two-layered works, i.e. Glimpse [13], use heuristics and lower level features to remove video redundancy, which fits better into current real-word deployment where cameras are usually cheap and the edge servers are not available.

Real time video analytics systems [13, 15, 16, 25, 27] optimize the whole pipeline, including camera processing delay, network overhead and server inference latency, to reduce end-to-end respond time for the inference tasks. For example, Reducto [27] assigns tiny workload to the cameras to avoid exaggerating camera processing delay, and hence, fulfill real time missions. Non-real-time systems [23, 30] try to answer "after the fact" types queries from large scale stored videos. The latter class of systems usually focus on the efficiency of key frame searching and high inference throughput.

The majority of video analytics systems are designed to process video streams independently [13, 14, 23, 27]. All optimization and redundancy pruning are within a single video stream, which leads to its linear growth resource requirements. The other systems [24, 25, 30] focus on the cross-camera analytics on a group of cameras. But they either fail to achieve real time inference, i.e. Caesar [30], or fails to provide comprehensive coverage to the surveillance scene, i.e. Spatula [25].

---
[1]Different applications may have different requirements to define a detection as effective. For example in a vehicle plate detection scenario, only the detection of the front or the back view is effective. In this paper, we assume an application that a detection from any viewpoint is sufficient to fulfill the mission. However, our system can be easily scaled to other scenarios given clear effectiveness definition, i.e., we only take the front/back views into our system in the vehicle plates detection scenario.

Different from all existing works, CrossRoI achieves real-time cross-camera video analytics over a fleet of cameras and fulfills comprehensive scene coverage. CrossRoI fits into a two-layered architecture which only assumes normal surveillance cameras without the needs of advanced edge devices.

## 2.2 Classic & Tile-based Video Compression

The vanilla video compression standard, i.e. H.264 AVC [43] and HEVC [40], are widely applied to significantly reduce data sizes in video storage/streaming applications. These compressors usually encode videos with two steps. (1) The encoders first split every frame into many small pixel *blocks* (for example, 16 *pixels* × 16 *pixels* block size for H.264 standard). For every block in a video frame, the compressor searches similar blocks either within the already-encoded portion of current frame or in nearby frames that are buffered by the encoder. When a closely matched block is identified, it encodes the position of this similar block in a motion vector. (2) The encoder calculates the pixels level difference between current block and the reference block, and encodes this sparse residual difference with quantization and entropy encoding. Video compression efficacy can be impacted by the frame size in such codecs. For example, a block has more reference block options when the frame size is large, and hence, more easily get encoded into space-efficient motion vector and sparse residual difference.

Tile-based video compression are widely used in data-intensive applications, for example, panoramic video streaming. [12, 17, 19, 44] Tile-based video encoder splits the whole frame spatially into several rectangular *tiles*. Every tile of the video is processed independently by a classic video compressor (i.e. H.264) and can be encoded into different qualities. For example, a compressor can encode the region-of-interest parts of a video with high bitrates and the other parts with low bitrates. While tile-based compression reduces video size through the semantic region-of-interest information, splitting a large video into several smaller ones degrades the overall efficacy of the compressors.

In CrossRoI, we applied tile-based video compression to include only the interesting regions of the surveillance videos. To alleviate the compression efficacy degradation, we design a tile grouping algorithms to merge small tiles into larger ones, which reduces network overhead even further compared to existing tile-based approaches (§4.3).

## 2.3 Computer Vision based Object Re-identification

CrossRoI establishes the cross-camera region associations through profiling object re-identification (ReID) results among the group of cameras. ReID is a challenging problem in computer vision [31, 38, 48]. A typical ReID pipeline starts with automatic object detection with object detectors, i.e. YOLO [34], FasterRCNN [37] and SSD [29]. ReID algorithms then extracts deep image features from the detected objects and computes the similarity of two detection based on their feature distance [21, 26, 29, 39]. Some works [21, 26] apply object movement patterns as spatial-temporal cues to further improve the identification accuracy. Although many ReID algorithms are proposed, the ReID results are still not perfect, especially in crowded scenes and large camera networks where ablations and significantly different lighting conditions and viewing angles are common. Different from computer vision communities, we do not reinvent new ReID algorithms in this paper, but apply statistical filters to augment existing ReID algorithms to obtain highly-confident region associations from error prone ReID results (§4.2).

## 3 CROSSROI BASIC MODELS, CONCEPTS AND PROBLEM DEFINITION

The CrossRoI system has two entities, which are *CrossRoI cameras* and *CrossRoI server*. CrossRoI cameras are the video providers. They capture videos of the mission scene and transmit them back to CrossRoI server for inference and analytics, e.g., car detection and counting. The workflow of CrossRoI contains an *offline phase* and an *online phase*. In offline phase, the CrossRoI server collects synchronized video clips from each CrossRoI camera. Through profiling and analyzing these clips, the server can calculate optimal RoI masks for the cameras. These RoI masks will then be applied in the online phase to reduce network overhead and boost inference throughput at the CrossRoI server. In the rest of this section, we focus on the offline video profiling process of CrossRoI to explore the following two questions:

- How to establish the data associations between multiple cameras covering the same scene?
- How to calculate the optimal RoI masks for these cameras without losing any interesting object?

Specifically, we first introduce the video data model in §3.1, and then answer the above two questions in §3.2 and §3.3, respectively. We will present more detailed system workflow and online phase designs in §4.

## 3.1 CrossRoI Data Model

We consider a CrossRoI system containing $N$ CrossRoI cameras, named $C_1, C_2, \ldots, C_N$. In the offline phase, the CrossRoI server collects synchronized video clips from all the CrossRoI cameras for profiling. "Synchronized" here refers that all $N$ video clips have the same frame rate $f$, start at the same time[2] $t_1$, as well as the same video length. The $k$-th frame of every video clip is then corresponding to the same timestamp $t_k = t_1 + \frac{k-1}{f}$, for any $k$ within the length of the videos. In this manner, the frames from CrossRoI cameras with the same indices are just image captures of the same scene at the same time from different perspectives. We further define the profile time window being a list of discrete timestamp $\mathcal{T} = \{t_1, t_2, \ldots, t_L\}$, where $t_i$ is the timestamp of the $i$-th frames in the video clips and $L$ is the index of the last frame.

In order to study the fine-grained data associations between cameras, we further cut every video into *tiles*. Tiles are smaller rectangular spatial regions which cumulatively cover the whole frame. As shown in Figure 2a, the whole frame area of $C_1$ is divided into 24 tiles indexed from 1 to 24 in an top-to-bottom, left-to-right order. We formally define $\mathcal{G}_i$ as the set of tiles for camera $C_i$, where $1 \leq i \leq N$.[3] The $j$-th tile of $C_i$ can then be referred as $\mathcal{G}_{i,j}$. For example, the left top tile of $C_1$ in Figure 2a is $\mathcal{G}_{1,1}$. It is worth mentioning that a tile is not corresponding to any specific frame or timestamp. Tiling is a spatial description of how we divide the field of views of cameras into finer granularity.

---
[2]We consider two cross-camera timestamps as the same if their difference is small enough for frame alignment, i.e. $< \frac{1}{2f}$, which can be achieved by NTP protocol.
[3]In this paper, we use "tiles of $C_i$" and "tiles of the video generated by $C_i$" interchangeably.

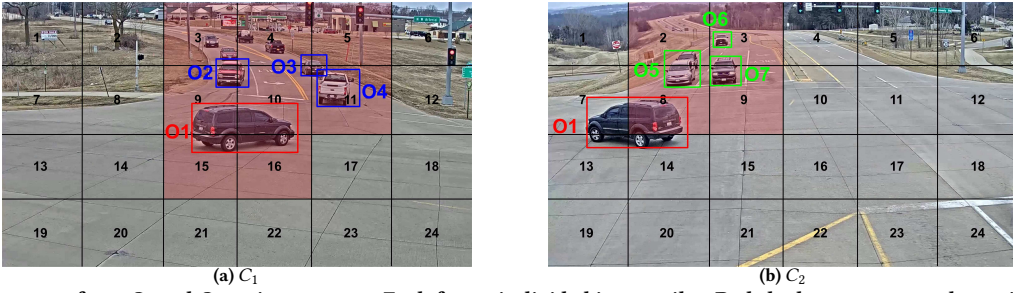

(a) $C_1$        (b) $C_2$

Figure 2: Video captures from $C_1$ and $C_2$ at timestamp $t_1$. Each frame is divided into 24 tiles. Red shadow represents the optimized RoI masks generated for these two cameras based on profiling on $t_1$ only.

| Timestamps | $t_1$ | ... |
|---|---|---|
| Detected Objects | $O_{t_1} = \{O_1, O_2, O_3, O_4, O_5, O_6, O_7\}$ | ... |
| Appearance Regions | $\mathcal{R}^1_{t_1} = \{\{\mathcal{G}_{1,9}, \mathcal{G}_{1,10}, \mathcal{G}_{1,15}, \mathcal{G}_{1,16}\},$ $\{\mathcal{G}_{2,7}, \mathcal{G}_{2,8}, \mathcal{G}_{2,13}, \mathcal{G}_{2,14}\}\}$ $\mathcal{R}^2_{t_1} = \{\{\mathcal{G}_{1,3}, \mathcal{G}_{1,4}, \mathcal{G}_{1,9}, \mathcal{G}_{1,10}\}\}$ $\mathcal{R}^3_{t_1} = \{\{\mathcal{G}_{1,4}, \mathcal{G}_{1,5}, \mathcal{G}_{1,10}, \mathcal{G}_{1,11}\}\}$ $\mathcal{R}^4_{t_1} = \{\{\mathcal{G}_{1,11}\}\}, \mathcal{R}^5_{t_1} = \{\{\mathcal{G}_{2,2}, \mathcal{G}_{2,8}\}\}$ $\mathcal{R}^6_{t_1} = \{\{\mathcal{G}_{2,3}\}\}, \mathcal{R}^7_{t_1} = \{\{\mathcal{G}_{2,3}, \mathcal{G}_{2,9}\}\}$ | ... |

Table 1: Cross-camera region association lookup-table for the two-camera example as shown in figure 2.

In CROSSRoI, we define *RoI mask* as the region in camera frames that may contain interesting objects, for example, vehicles or trucks. The regions outside of a RoI mask are ignored in the video analytics pipeline because no interesting targets may appear in these areas. In our system, a tile is the smallest spatial unit to constitute a RoI mask. The RoI mask for camera $C_i$, denoted as $\mathcal{M}_i$, is a subset of all its tiles, i.e., $\mathcal{M}_i \subset \mathcal{G}_i$. For example in figure 2a, if we want RoI region to only include the four detected cars, then the minimum-sized RoI mask will be as follows.

$$\mathcal{M}_1 = \{\mathcal{G}_{1,3}, \mathcal{G}_{1,4}, \mathcal{G}_{1,5}, \mathcal{G}_{1,9}, \mathcal{G}_{1,10}, \mathcal{G}_{1,11}, \mathcal{G}_{1,15}, \mathcal{G}_{1,16}\}$$

Assigning smallest RoI masks to each camera without losing detection of interesting objects is a non-trivial task, especially when we want the masks fit long profiling duration. We will present our RoI masks generation scheme in §3.3.

### 3.2 Cross-Camera Regions-Association Concept

We establish cross-camera region associations based on existing object re-identification (ReID) algorithms, which take visual or geographical features to associate common objects across multiple frames/ cameras. ReID algorithms assign an ID for every detected object. Detection of the same object across different cameras will be assigned to the same ID. For example, Figure 2a and Figure 2b are two synchronized frames from $C_1$ and $C_2$, respectively. Every detected car in both frames is assigned an ID by the ReID algorithm. Cars $O_2, O_3$, and $O_4$ are unique to $C_1$. Cars $O_5, O_6$, and $O_7$ are unique to $C_2$. While car $O_1$ appears at the overlapping region of both views and can be identified in $C_1$ and $C_2$ simultaneously.[4]

We establish the cross cameras association by profiling the ReID results over the whole time window $\mathcal{T}$. At any timestamp $t_m$, we record the following two elements:

- All objects being detected at this timestamp, denoted as $O_{t_m}$,
- The *appearance regions* for each object being detected at this timestamp.

We define the *appearance region* of an object $O_k$ on camera $C_i$ at timestamp $t_m$ as the least set of tiles that can cover $O_k$, denoted as $R^k_{i,t_m}$. As the object may appear on multiple cameras simultaneously, we further define the *appearance regions* of $O_k$ at timestamp $t_m$, denoted as $\mathcal{R}^k_{t_m}$, as the collection of its appearance regions over all CROSSRoI cameras, s.t.,

$$\mathcal{R}^k_{t_m} = \{R^k_{i,t_m} | 1 \leq i \leq N \text{ and } R^k_{i,t_m} \neq \emptyset\}$$

Take figure 2 as an example, There are seven objects being detected at $t_1$ in total, s.t., $O_{t_1} = \{O_1, O_2, O_3, O_4, O_5, O_6, O_7\}$. $O_1$ appears in both frames, therefore its appearance regions contains two elements, s.t.,

$$\mathcal{R}^1_{t_1} = \{\{\mathcal{G}_{1,9}, \mathcal{G}_{1,10}, \mathcal{G}_{1,15}, \mathcal{G}_{1,16}\}, \{\mathcal{G}_{2,7}, \mathcal{G}_{2,8}, \mathcal{G}_{2,13}, \mathcal{G}_{2,14}\}\}$$

The other objects appear only once and thus have single-length appearance regions, e.g. $\mathcal{R}^5_{t_1} = \{\{\mathcal{G}_{2,2}, \mathcal{G}_{2,8}\}\}$. Profiling through the whole time window, we can build a lookup-table which ensembles the ReID based region associations, as shown in table 1. We will show how to generate optimal RoI masks with the region associations in §3.3 shortly.

As mentioned in §2, ReID algorithms are still not perfect. In order to achieve accurate region associations based on the error prone ReID results, we apply statistical filters on the raw ReID results to obtain highly-confident ReID results and establish the region association with the selected data instead. We will show more details about the filter design in §4.2.

### 3.3 RoI Masks Optimization

The optimization objective is to *include least number of tiles into the RoI masks cumulatively across all the N cameras*. In order to avoid missing any object at any timestamp in the time window, we set the optimization constraints as *any object occurred at timestamp $t_m$ has at least one appearance region included by the RoI masks, for*

---
[4]We only show detection of large objects in the two frames in figure 2 for clarity of illustration.

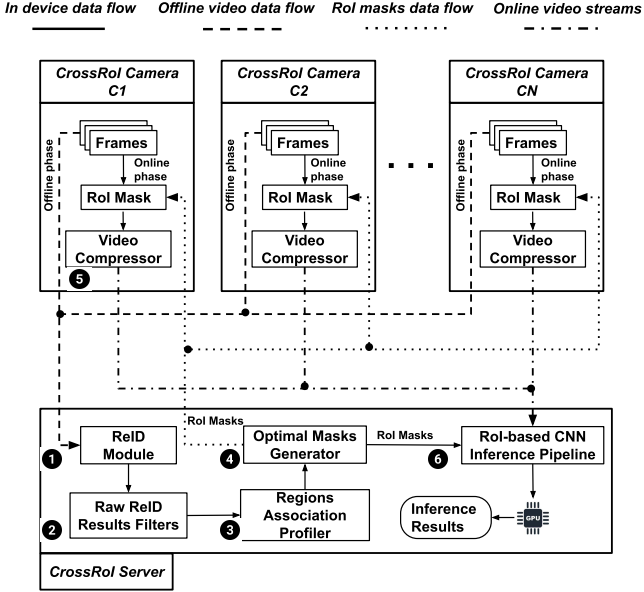

Figure 3: System Overview of CrossRoI

*any* $t_m \in \mathcal{T}$. We define variable $\mathcal{M}$ as the union set of all the $N$ RoI masks, s.t. $\mathcal{M} = \cup_{i=1}^{N} \mathcal{M}_i \subset \cup_{i=1}^{N} \mathcal{G}_i$. The optimization problem can be formally presented as follow:

$$\min \ |\mathcal{M}| \tag{1}$$

$$\text{s.t.} \sum_{R \in \mathcal{R}_{t_m}^k} \left( \mathbb{1}(R \in \mathcal{M}) \right) \geq 1, \quad \forall \ t_m \in \mathcal{T}, \ \forall \ k, \ \text{s.t.,} \ O_k \in O_{t_m} \tag{2}$$

Note that $\mathbb{1}(\cdot)$ in equation 2 is an indicator function with its range being $\{0, 1\}$. It will return 1 if the input condition, e.g., $R \in \mathcal{M}$ in equation 2, is true and return 0 otherwise. Solving the above combinatorial optimization will generate the optimal RoI masks which include least number of tiles while ensuring every object being detected. In the example of figure 2, if we set the time window to include $t_1$ only, the optimized RoI masks $\mathcal{M}$ will be

$$\{\mathcal{G}_{1,3}, \mathcal{G}_{1,4}, \mathcal{G}_{1,5}, \mathcal{G}_{1,9}, \mathcal{G}_{1,10}, \mathcal{G}_{1,11}, \mathcal{G}_{1,15}, \mathcal{G}_{1,16}, \mathcal{G}_{2,2}, \mathcal{G}_{2,3}, \mathcal{G}_{2,8}, \mathcal{G}_{2,9}\}$$

, which are shown in the figures with pink shadow. All the appearance of $O_2, \ldots, O_7$ are covered by the RoI masks. As $O_1$ appears simultaneously on both cameras, the algorithms will only include one of its appearance regions that introduce least overheads, e.g., its appearance region in $C_1$ in this example. The optimized RoI masks will then be applied in the online phase to boost overall system performance (§4).

## 4 DESIGN AND IMPLEMENTATION
### 4.1 System Overview
As mentioned in §3, the CrossRoI system has an offline phase and an online phase. In offline phase, the server generates optimal RoI masks through profiling synchronized video clips. In online phase, the server runs video analytics tasks in real time, where the RoI masks serve as a guidance to reduce network burden and boost server throughput. Figure 3 depicts the high-level framework of CrossRoI. We show more details of the workflow as follows.

*4.1.1* **Offline Phase.** *Offline server re-identification* ❶. The CrossRoI server first applies re-identification algorithms over several minutes of synchronized raw videos collected from the CrossRoI cameras to characterize the view relations among different cameras. We choose DiDi-MTMC [26] algorithm as the CrossRoI server ReID module, which integrates vision features together with geographic information to achieve best accuracy on our experiment dataset (dataset details are presented in §5). At the end of the ReID step, every interesting object (vehicles in our scenario) of every frame is associated with a bounding box and an ID, in the form of *<left, top, width, height, id>*. *Left* and *top* information locate the top left corner of the bounding box, while *width* and *height* information characterize the bounding box size. All four values are measured in terms of pixel(s). These ReID results will be further processed in modules ❷, ❸ and finally be used to generate the optimized RoI masks.

*Raw ReID Results Filtering* ❷. Although the ReID module achieves state-of-the-art accuracy, its results still contain a lot of errors and mismatches. In order to establish accurate region associations among the CrossRoI cameras, we pass the ReID results through two tandem statistical filters to remove the "suspicious" ID assignments and only keep highly-confident ReID results. At the end of this step, we get a selected set of highly-confident ReID results, which will be used in module ❸ for cross camera region associations. It is worth mentioning that the two filters do not improve overall ReID accuracy compared to existing ReID algorithms. The design goal of statistical filters is to select a subset of highly-confident object identification results, which are effective to represent the cross-camera associations.

*Regions association & RoI masks generation* ❸, ❹. Based on the filtered ReID results obtained in step ❷, the server builds a lookup table, e.g. Table 1, that ensembles the region associations across all the cameras. The CrossRoI server then takes the table as input into the optimization framework and generates optimized RoI mask for every camera, as described in §3. We use commercial optimization solver (i.e. Gurobi [7]) to obtain the optimization results in the offline phase. At the end of the offline phase, CrossRoI server sends the corresponding RoI mask to each CrossRoI camera. Hence, the cameras can use RoI masks to crop and further compress their video streams in online phase, as shown in ❺. The CrossRoI server will also keep the RoI masks in memory and applies them onto the CNN inference tasks to boost its execution speed, as shown in ❻.

*4.1.2* **Online Phase.** *Oline phase video compression and streaming* ❺. In the online phase, CrossRoI cameras stream their video feeds to the server in real time. The server runs CNN based inference algorithm on these videos to answer the queries (e.g. vehicles detection or counting). In order to reduce server side bandwidth consumption caused from receiving many video streams at the same time, all CrossRoI cameras (1) crop their videos and only stream the areas included by RoI masks, and (2) apply modern video compressor (e.g., H.264) to greatly reduce the video size. To boost video compression efficacy on cropped videos, we develop a *tile grouping* algorithm based on H.264 codec which merges the fine-grained small tiles into larger ones, and hence, further reduces the videos size. We present more details about this algorithm in §4.3.

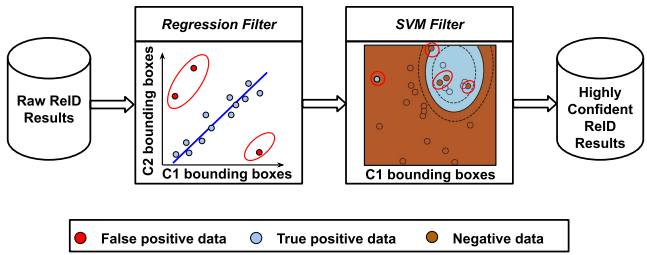

Figure 4: Statistical filters for raw ReID results. Outliers are circled out in red color.

*RoI based real-time CNN inference* ❻. Once the tile streams are received by the CROSSROI server, they will be merged together to reconstruct frames. Note that the non-RoI regions of a frame will be empty (purely black) as the corresponding tiles are not streamed to the server. These recovered frames are then pushed into a RoI based CNN inference pipeline. In our system, we choose YOLO [34] as the inference handler for object detection task. Different from traditional object detection tasks, where an interesting target may appear everywhere in the whole frame, a RoI based object detection task has prior knowledge of the RoI regions, and thus, can greatly reduce the detection space (i.e. only run YOLO on the RoI regions). In CROSSROI system, we build a RoI-YOLO detector based on SBNet [36] which takes the RoI masks as cues to boost YOLO detection speed by 1.2x. More details are presented in §4.4.

## 4.2 Raw RoI Results Analysis and Filtering (Offline)

*4.2.1 Raw ReID Results Analysis.* . We first present a comprehensive analysis towards the raw ReID results, which sheds light to our design of the two tandem statistical filters to remove "suspicious" ReID data points. We investigate pairwise ReID results between two different cameras to understand the structure of these raw results and where mistakes happen. In the study of a pair of cameras, we categorize any ID assignment of a *source camera*, into one of the two types, *positive* and *negative*, in terms of whether the detected object has an appearance in the *destination camera* at the same timestamp.

To better illustrate the concepts, we use $C_1$ and $C_2$ in figure 2 as an example pair of cameras in the rest of this subsection. We set $C_1$ as the source camera and $C_2$ as the destination camera. Every object being detected in this frame will be assigned a positive or negative label, i.e. $O_1$ is positive and $O_2, O_3, O_4$ are negative. In our example, every object is identified correctly. However, the ReID algorithm may make mistakes, e.g. assigning same ID to two different objects or assigning different IDs to multiple appearances of the same one. To illustrate the correctness of the identification, we further associate a correctness label, being either *true* or *false*, to each detected object in the source camera. Hence, there are four types of labels associated with all the identification results as follows:

- **True Positive (TP)**. A true positive label is assigned to an object in the source camera which has a corresponding appearance in the destination cameras, and these two appearances are given the same ID, e.g., $O_1$ in $C_1$ is a true positive data point.
- **False Positive (FP)**. A false positive label covers either of the following two cases: (1) a negative object being matched to an object in the destination camera, i.e. in case $O_2$ in $C_1$ and $O_5$ in $C_2$ being assigned the same ID, and (2) a positive object being matched to a wrong object in the destination camera, e.g., in case $O_1$ in $C_1$ being matched $O_7$ in $C_2$.
- **True Negative (TN)**. A true negative label refers to an object which has no appearance in the destination camera and that its identification is correct, e.g., $O_2, O_3, O_4$ in $C_1$.
- **False Negative (FN)**. A false negative label is corresponding to the case when a positive object being mistakenly identified as a negative object. For example, in case the ReID algorithms fails to find the appearance of $O_1$ in $C_2$ and assign different IDs to these two appearance of the same object.

Both false data (FP and FN) will sabotage the optimized RoI generation framework as mentioned in §3. Specifically, (1) the false positive data will make the generated RoI masks incorrect, as it introduces wrong region associations between cameras, and (2) the false negative will significantly degrade the efficacy of non-RoI tiles reduction, as we try to ensure every object has a least one appearance at any time. For example, if $O_1$ in $C_1$ and $C_2$ are assigned different IDs at $t_1$, we must include both $\{\mathcal{G}_{1,9}, \mathcal{G}_{1,10}, \mathcal{G}_{1,15}, \mathcal{G}_{1,16}\}$ and $\{\mathcal{G}_{2,7}, \mathcal{G}_{2,8}, \mathcal{G}_{2,13}, \mathcal{G}_{2,14}\}$ into the RoI masks forever no matter what identification happens in later timestamps.

To better understand the distribution of the above four types of ReID results. We profile a dataset containing synchronized videos from five traffic cameras watching the same crossing (we will describe more details about the dataset in §5). We compare DiDi-MTMC ReID algorithm to the ReID ground truth of the dataset and get the distributions of the pairwise ReID results as shown in table 2. It can be observed that there are large amount of falsely identified cases, especially the false negative identifications which usually outweigh the total number of true/false positive samples. Applying raw ReID results to the optimization pipeline will definitely lose many optimization opportunities and degrade the system efficacy.

Although the raw ReID results are error prone, we make two important observations after close scrutiny of the application scenario and results distribution (Table 2), which can help remove the false ReID results significantly. The two **observations** are as follows:

(O1) The region-associations between two cameras have intrinsic physical relation. For example, the two appearances of $O_1$ at $t_1$ suggest that region $\{\mathcal{G}_{1,9}, \mathcal{G}_{1,10}, \mathcal{G}_{1,15}, \mathcal{G}_{1,16}\}$ in $C_1$ and region $\{\mathcal{G}_{2,7}, \mathcal{G}_{2,8}, \mathcal{G}_{2,13}, \mathcal{G}_{2,14}\}$ in $C_2$ are actually the same area in physical means. In any future frames, this mapping relation will also work.
(O2) In both positive and negative identifications of the ReID results, the number of true samples is always greater than that of false samples, and usually greater in several times or magnitudes.

Based on the above two observations, we decide to apply statistical filters to remove the false ReID results. Specifically, we design a *regression filter* to remove the false positive samples and a *SVM filter* to remove the false negative samples. More details will be presented in §4.2.2 and §4.2.3.

| S \ D | $C_1$ TP | FP | FN | TN | $C_2$ TP | FP | FN | TN | $C_3$ TP | FP | FN | TN | $C_4$ TP | FP | FN | TN | $C_5$ TP | FP | FN | TN |
|---|---|---|---|---|---|---|---|---|---|---|---|---|---|---|---|---|---|---|---|---|
| $C_1$ | | | | | 335 | 253 | 263 | 7542 | 358 | 22 | 560 | 7453 | 162 | 15 | 336 | 7880 | 101 | 0 | 642 | 7650 |
| $C_2$ | 333 | 253 | 291 | 4317 | | | | | 161 | 81 | 397 | 4551 | 242 | 56 | 401 | 4497 | 50 | 2 | 773 | 4371 |
| $C_3$ | 358 | 22 | 977 | 8246 | 161 | 81 | 868 | 8558 | | | | | 434 | 40 | 951 | 8243 | 155 | 24 | 1871 | 7618 |
| $C_4$ | 162 | 15 | 512 | 6784 | 242 | 56 | 917 | 6258 | 434 | 40 | 809 | 6190 | | | | | 138 | 22 | 1402 | 8583 |
| $C_5$ | 101 | 0 | 694 | 8568 | 50 | 2 | 1074 | 8237 | 155 | 24 | 1552 | 7632 | 138 | 22 | 1328 | 7875 | | | | |

**Table 2: Characterization of raw ReID results. S/D represents the source/destination camera. For each pair of cameras, we count the number of identifications with four different *labels*, which are TP, FP, FN, TN, representing *true positive, false positive, false negative* and *true negative*, respectively. Detail descriptions for each label type are presented in §4.2.1.**

*4.2.2* **Regression Filter Design and Implementation.** As shown in Figure 4, we push raw ReID data through two tandem filters to get cleaned. The first filter is a regression filter. We dump all the positive results into a regression module to learn the intrinsic region mappings between a pair of cameras. We use regression method here for its reliable and successful applications to model correlations between a pair of dependent variables, e.g., appearances of same objects in source/destination cameras. The outliers of the trained model are regarded as false positive samples and will be rectified.

Specifically, we feed the two bounding boxes of a positive object in its source camera and its destination camera to the regression function, i.e. $<bbox^1_{C_1}, bbox^1_{C_2}>$ for our example at $t_1$, where $bbox^1_{C_1}$ represents the bounding box of $O_1$ from $C_1$ and $bbox^1_{C_2}$ represents that from $C_2$. All the bounding boxes are 4D vectors in the form of *<left, top, width, height>*. We apply regression filter mechanism based on our observation that similarly localized bounding boxes with similar sizes are objects at the same physical locations and their corresponding appearances on the destination camera should also be homogeneous. An outlier of the regression results is very likely to be a false positive sample. After the regression filter, we get a subset of positive data outliers and regard them as the false positive samples. Instead of directly removing these data, we choose to decouple the incorrect association between its counterpart in destination by assign it a new ID. This data point will then be regarded as a negative data sample to go through the SVM filter.

In our system implementation, we use the robust regression module of `sklearn` [9] as our regressor. As the mapping relation between two cameras may not be simply linear, we apply higher order features of the data to make the filter fit ReID results better. Specifically, we use RANSAC [18] algorithm as the kernel algorithms of regression as its regression process naturally splits data samples into inliers and outliers, and hence, fits the purpose of our regression filter design. We fine-tune its `residual-threshold` parameter, which determines threshold distance for a sample to be regarded as an outlier, to find the best performance. We will show more evaluations about our filter mechanism in §5.

*4.2.3* **SVM Filter Design and Implementation.** After the regression filter, we push all the raw ReID data, both positive and negative samples, into the SVM [22] filter. In this step, we want to learn an accurate two-class clustering between positive and negative ReID data samples based on their position-and-shape features (i.e., bounding box position and size). We choose SVM as the second step filtering model for its widely successful application in two/multiple class classification.

In our case, we feed positive data to SVM in the form of *<bbox, 1>* and negative data in form of *<bbox, 0>*. We push all data samples into SVM to train a model and apply this model back to the ReID data to obtain outliers. It is worth mentioning SVMs are usually trained and tested with different data. However, we train and use the SVM model on the same data because we are not generating a classifier for future data but applying it as a filter on existing samples. We fine tune hyper parameters in SVM to avoid model overfitting, and hence, generate no outliers. The outliers here refer to negative samples appeared in positive regions and positive samples in negative regions, as shown in Figure 4. As we have much less positive data and have already removed positive outliers in the regression filter, we do not further remove positive outliers in SVM filter. We regard the negative outliers as false negative samples and directly remove these data from entering the optimization process. We choose to remove false negative samples only because (1) it is impossible to correctly make this sample "positive" by locating its counter part in destination camera, which is not achieved even by the state-of-the-art ReID algorithms, and (2) due to the redundancy of region associations, i.e. different objects at different timestamp usually convey the same regions mapping, the region associations usually do not change without several pairs of data samples. At the end of SVM filtering, we remove the false negative data samples. The remaining ReID results are highly confident and will go through the profiling and optimization framework in ❸ and ❹.

In our system implementation, we use the SVM module with of `sklearn` [9] as our filter. We fine-tune its $\gamma$ parameter, which determines the SVM kernel non-linearity, to explore the best performance. More evaluations about SVM filtering are presented in §5.

*4.2.4* **Discussion.** As both regression filter and SVM filter are statistical, it is impossible to ensure the filtering is perfectly accurate. It is possible that we can not remove all the false identification. The filtering mechanism may even remove true identification results, either true positive or true negative. However, due to the redundancy of region associations, especially when we profile through videos long enough (containing thousands of frames), the CrossRoI system accuracy will not be degraded by the harsh filtering, while the system efficacy gets boosted significantly. We will show more CrossRoI system evaluations in §5.

|  | *original* | 2 × 2 | 2 × 4 | 4 × 4 | 4 × 8 | 8 × 8 |
|---|---|---|---|---|---|---|
| $C_1$ | 82.7 | 85.9 | 86.2 | 89.0 | 90.4 | 97.3 |
|  | **(1)** | **(1.03)** | **(1.04)** | **(1.07)** | **(1.09)** | **(1.17)** |
| $C_2$ | 121.2 | 124.5 | 124.8 | 127.6 | 129.6 | 136.2 |
|  | **(1)** | **(1.03)** | **(1.03)** | **(1.05)** | **(1.07)** | **(1.12)** |
| $C_3$ | 102.2 | 103.3 | 103.6 | 105.2 | 106.4 | 112.9 |
|  | **(1)** | **(1.01)** | **(1.01)** | **(1.03)** | **(1.04)** | **(1.10)** |
| $C_4$ | 97.9 | 99.3 | 99.5 | 100.0 | 101.7 | 108.6 |
|  | **(1)** | **(1.01)** | **(1.01)** | **(1.02)** | **(1.04)** | **(1.11)** |
| $C_5$ | 40.9 | 41.1 | 41.4 | 42.0 | 43.2 | 47.4 |
|  | **(1)** | **(1.01)** | **(1.01)** | **(1.03)** | **(1.06)** | **(1.16)** |

**Table 3: Efficacy characterization of tile-based video compression. Videos are either compressed with original H.264 standard or split into $m \times n$ tiles (e.g., $2 \times 4$) and compressed with tile-based method accordingly. Video-sizes are measured in unit of MB. Bold numbers represent the video size amplifications compared to *original* video compression without tiling.**

### 4.3 Tile Based Video Compression and Streaming (Online)

**4.3.1 Characterizing Tile-based Video Compression.** In online phase, the CrossRoI cameras apply RoI masks on their video captures to crop the videos and remove all the non-RoI tiles. The tiles included in RoI masks will be further compressed by video compressors to reduce its file size before being streamed over the network, as shown in ❺. However, applying video compressor on each tile of video separately greatly degrade the efficacy of modern video compressors, e.g. H.264. As mentioned in §2, compressors reduce video size by exploring the content similarity among existing blocks, cutting videos into small tiles reduces the number of references each block may refer to and thus degrades compression efficacy. To better illustrate the performance degradation, we profile on our dataset (§5) by cutting five different videos into different-sized tiles and encoding them in H.264 format to characterize the compression efficacy of the video compressor. As shown in Table 3, we split the videos according to five settings, each split the videos into $m \times n$ tiles evenly (e.g. $2 \times 4$). As we split the video in finer-grained, the total video sizes grow larger, which indicates a degradation of video compression efficacy.

**4.3.2 Tile Grouping Algorithm.** In order to improve video compression efficacy, we develop a straight-forward greedy-based *tile grouping* algorithm to merge fine-grained small tiles in RoIs masks into larger ones to further reduce the video-sizes being sent to the CrossRoI server over network. As shown in Figure 5(a), the video is cut into 6 × 5 small tiles. The white tiles are included in the RoI mask, while the shadow tiles are in non-RoI region. The tile grouping algorithm interactively find the largest inscribed rectangular in the RoI masks and merge all small tiles in this rectangle into a large tile until every tile in RoI mask is processed. For example, in figure 5(b), we first merge all the 12 tiles covering region 1 into a large tile, and then merge the remaining 4 tiles into two large tiles, respectively. In this way, we merge the original 16 small tiles into 3 larger ones, and hence, improve the compression efficacy.

Finding largest inscribed rectangular in a binary grid can be easily solved with dynamic programming in $O(M)$ time, where $M$ is the number of small tiles in the video. The overall time complexity of the tile grouping is hence upper bounded by $O(M^2)$. Furthermore, the tile grouping results can be calculated in offline phase once the RoI masks are generated. Therefore, the tile grouping algorithm will introduce zero overhead to the CrossRoI cameras in online phase. It is worth mentioning that our tile grouping algorithm is a heuristic greedy algorithm which cannot ensure the generated groups is exactly the optimal way to merger tiles. However, we show the significant improvement of video compression efficacy when applying our algorithm through experimental evaluations in §5.

**4.3.3 Implementation.** In CrossRoI cameras, we choose ffmpeg [6] H.264 implementation as our video compressor. The video compressor will queue a segment of video frames, i.e. 2s or 20 frames, and compress these images into a short video before sending it to the server. A longer segment benefits video compression efficacy, as the more temporal redundancy can be reduced, but increases server response delay for detecting objects in video. We will show more evaluations on video segment length in §5.

### 4.4 RoI Based CNN Inference (Online)

Once the CrossRoI server receives video feeds from the cameras, it will dump these videos into the video analytics pipeline, which loads both video data and CNN-based machine inference models (e.g. YOLO object detector) to GPU and finally return the detection results (e.g. bounding boxes of vehicles) ❻. Traditional CNN models usually have a respective field of the whole frame, which is not optimized for our case where the prior knowledge of RoI masks is available. In CrossRoI server, we prefer a RoI-based inference pipeline, where the CNN model works on the RoI covered data only, and hence, boosts the system inference speed.

In the CrossRoI server, we choose to implement the RoI-based CNN inference pipeline based on SBNet [36], which is an optimized CUDA kernel specially designed for RoI based CNN inference tasks. Image data is usually transformed into 4D tensors, in the form of *<batch, height, width, channel>*, when being processed in GPU. SBNet divides the input tensor into small tiles in the *height* and *width* dimensions. It gathers all the RoI "tensor-tiles" and stacks them together to generate a small and deep tensor constituting of RoI-covered data only. As presented in figure 6, SBNet kernel adds a *gather* module before each convolutional layer of a CNN model to generate the "RoI tensor". SBNet then passes the new tensor through the convolutional module to get the data manipulated by the model. After the convolutional layer, SBNet adds another *scatter* module to transform the narrow tensor back to the original shape.

Based on SBNet, we build a RoI-YOLO object detector with Tensorflow [10]. It is worth mentioning that although SBNet can boost the system inference speed significantly (i.e. 1.5 ∼ 2.5×) when the RoI area is small (10% ∼ 20%) compared to the whole frame, SBNet introduces computational overhead (i.e. gather and scatter) compared to traditional CUDA kernel and may not perform as well when the RoI area is close to the whole frame. In practice, we load both RoI-YOLO and normal YOLO into GPU and push large RoI-area videos to normal YOLO model instead to achieve best performance. More evaluations about our CNN inference model will be presented in §5.

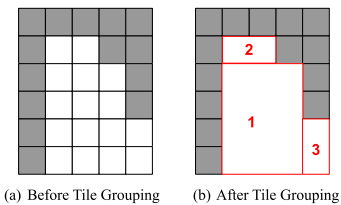

(a) Before Tile Grouping    (b) After Tile Grouping

**Figure 5:** *Tile grouping algorithm.* White tiles are corresponding to the RoI mask regions. Shadow tiles are out of the RoI mask.

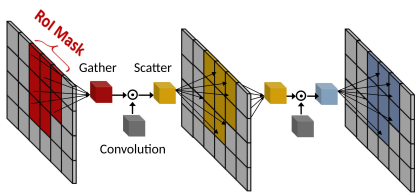

**Figure 6:** SBNet architecture illustrated with the RoI mask as shown in figure 2a. This figure is modified based on the SBNet paper [36].

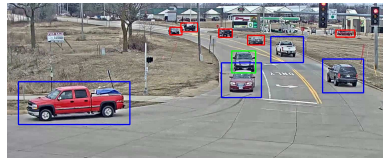

**Figure 7:** ReID ground truth augmentation. *Blue bboxes* represent original detection provided by the dataset ground truth. The *green bbox* is missing in the ReID ground truth due to ablation. *Red bboxes* are not included in original ground truth because they are out of the view-overlapping region of the cameras.

## 5 EVALUATION

### 5.1 Methodology

*5.1.1 Dataset.* We evaluate our system with *AI City Challenge 2020* traffic video dataset published by NVIDIA [32]. The dataset consists of two types of scenes where the traffic cameras are deployed either along long streets or around a traffic intersection, in a northern American city. We choose the most challenging scene of type two to evaluate CrossRoI, where 5 cameras are deployed around a traffic crossing with complicated inter-camera viewpoint overlapping. We present the locations and viewing angles of the five cameras in Figure 1. The dataset provides 5 synchronized videos taken from five cameras with 10 fps frame rate. The length of the videos ranges from 193 ∼ 215 seconds. We choose their overlapped 180s to evaluate the CrossRoI system. All the five videos have 1920 *pixels* × 1080 *pixels* resolutions (1080p) except the video generated by $C_5$, which is 1280 *pixels* × 960 *pixels*.

The five videos in scene 1 capture more than 30K vehicle bounding boxes over 3 minutes. Ground truth for vehicle re-identification (ReID) is provided with the dataset. However, the ReID ground truth has a shortcoming that it is very sensitive to occlusion, i.e. when vehicle *A* occludes vehicle *B* slightly, the ground truth will miss the detection and identification of *B*, while *B* could actually be detected by object detectors clearly. This usually leads to the "disappearance" of a vehicle for several frames, when it is partially occluded by other cars, in its continuous occurrence over the scene. Hence, we apply Kalman filter to fill the disappearance gaps in vehicles consecutive appearance. Another shortcoming of the ReID ground truth is that it only detects vehicles passing through multiple cameras and misses those vehicles appearing in a single camera only. We solve this limitation by augmenting the ReID ground truth with YOLO object detection results and assign unique ids to the vehicles not originally included in ReID ground truth. Figure 7 shows an illustrative example of our ground truth augmentation method.

*5.1.2 Evaluation Scenario & Metrics.* In our evaluation, we consider the query scenario as *unique vehicle detection*. Specifically, we want to detect every unique vehicle across all cameras at the scene in real time. As shown by the example in figure 2, there are 7 vehicles across the scene covered by $C_1$ and $C_2$ with 8 appearance bounding boxes. Our query scenario requires at least one detection bounding box for each unique object. Therefore, reporting either one of the two bounding boxes of $O_1$ fulfills the query requirement. As the CrossRoI system has two phases, we apply first 60s of the five videos as the input of offline phase to generate the RoI masks and evaluate the online phase system performance with the last 120s. The performance evaluation consists of the following four metrics:

(1) **Results Accuracy**. We define accuracy error as the absolute value of the percentile difference on the number of detected unique cars between the correct and system returned value. Hence, the accuracy is defined as one minus the error. As the dataset does not provide ground truth for vehicle detection, we fuse the ReID ground truth and raw YOLO detection results as the correct reference in our evaluation.
(2) **Network Overhead**. We define network overhead as the average bandwidth usage for CrossRoI server to download online video feeds from the cameras in real time.
(3) **System Throughput**. We define the system throughput as two parts: (1) the speed for the CrossRoI server to run vehicle detection inference in the online phase, and (2) the speed for CrossRoI cameras to compress video streams in real time.
(4) **End-to-end Respond Latency**. The average delay for CrossRoI server to generate vehicle detection results in the online phase. This latency includes camera side processing delay, networked latency and CrossRoI server processing overhead.

*5.1.3 Hardware & Implementation.* We deploy CrossRoI service on a server with 2 GeForce RTX 2080 Graphics Card, each with 2944 CUDA cores. The server has an Intel i7-9700K 8-core CPU and 64GB memory. The CrossRoI cameras are emulated on a laptop computer with an Intel i7-8850H 6-core CPU with 16GB memory. The laptop achieves 23 fps throughput for H.264 video compression on 1080p videos. Its performance is similar to most surveillance cameras which can achieve 25 ∼ 30 fps throughput on 1080p videos (e.g. Arecont Vision MegaVideo [3] and Logitech C930e [8]). The recorded videos are stored onto the laptop and streamed out to the server in real time with ffmpeg. The cameras and server are connected with emulated WiFi networks of 30 Mbps bandwidth and 10 ms round-trip-time. In this evaluation, we choose 64 *pixels* × 64 *pixels* as basic tile size to constitute the RoI masks for all five cameras.

### 5.2 Ablation Studies

We compare CrossRoI with four alternative methods to verify its merits and some of our design choices. Each alternative achieves "partial" functionality of CrossRoI by turning off one or some of

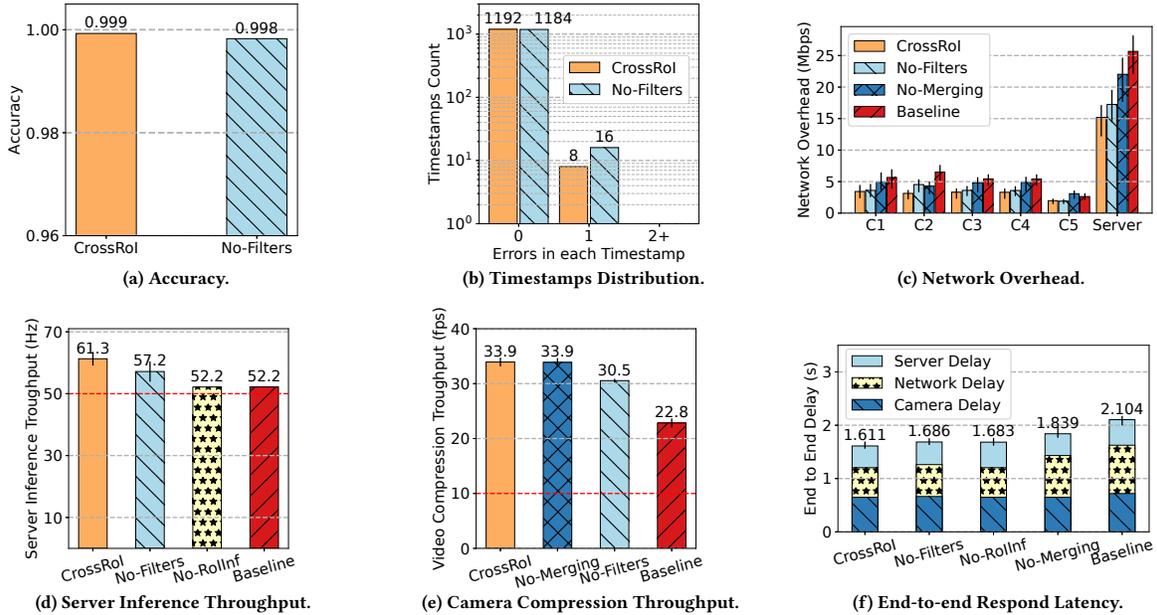

Figure 8: Performance Evaluations between CRossRoI and alternative methods.

CrossRoI's functional modules. The details of the alternatives are as follows.

(1) **Baseline**: *All CrossRoI functions* are turned off. Video streams are compressed with original H.264 compressor. The server runs off-the-shelve YOLOv3 [35] model as object detector to handle vehicle detection queries.
(2) **No-Filters**: *Regression & SVM Filters* ❷ are turned off, but the other modules remain. Raw ReID results are dumped into the RoI masks generation framework. Cameras crop their video streams in online phase based on the corresponding RoI masks.
(3) **No-Merging**: *Tile Grouping Algorithm* ❺ is turned off, but the other modules remain. The cameras compress their video streams into fine-grained tiles without merging them into larger ones.
(4) **No-RoIInf**: *RoI-based CNN inference* ❻ is turned off, but the other modules remain. RoI-YOLO model is replaced with off-the-shelve YOLOv3 model in the GPU inference step ❻.

The evaluation results between CrossRoI and the four alternatives are shown in figure 8.

*5.2.1 Accuracy.* As the dataset does not provide ground truth for vehicle detection, we set the detection results generated by *Baseline* method as the correct reference and fuse it with ReID ground truth to obtain the correct baseline for *unique vehicle detection* task. The *Baseline* method achieves 100% accuracy naturally as it sends full video data. We present the accuracy achieved by CrossRoI and *No-Filters* in figure 8a. CrossRoI achieves an accuracy of 99.9% that only 8 vehicles are missed in total 15424 vehicle appearances over 1200 timestamps. We plot the distribution of the timestamps in figure 8b in terms of how many vehicles are missed at each timestamp. It is easily observed that CrossRoI achieves correct detection for 1192 timestamps over the two minutes interval. There is at most one vehicle missed in the other 8 timestamps. The accuracy of *No-Filters* method is 99.8%. CrossRoI achieves a higher accuracy than *No-Filters* with less video data because the regression filter rectifies false positive associations in raw ReID results and improves the overall accuracy.

*5.2.2 Network Overhead.* We present network overhead for each camera and server in figure 8c. CrossRoI consumes least bandwidth compared to all other alternatives. The overhead of CrossRoI (15.2 Mbps) is by 42% reduced compared to *Baseline* method (26.2 Mbps). Comparing with *No-Filters* (16.5 Mbps), CrossRoI reduces more video redundancy by applying the SVM filter, which removes false negative samples in raw ReID results and generates smaller-sized RoI masks. CrossRoI reduces 30% network overhead compared to *No-Merging* method due to applying tile grouping algorithm to further improve the video compression efficacy.

*5.2.3 System Throughput.* We present inference throughput of the server in figure 8d and camera video compression throughput in figure 8e. The red lines represent the the minimum requirements for real time execution. That is, the server inference speed needs to be at least 50 Hz and the camera H.264 encoding throughput should be no less than 10 fps.[5] It can be observed that CrossRoI achieves highest throughput on both server (61.3 Hz) and camera (33.9 fps) sides. The RoI-based YOLO model improves overall server inference throughput by 18%. Compared with *No-Merging* method (33.9 fps), CrossRoI improves compression efficacy (i.e., reducing video sizes) without degrading compression processing speed.

*5.2.4 End-to-end Respond latency.* As in figure 8f, CrossRoI generates least end-to-end response delay (1.61 s) comparing to all the other alternatives. Compared with the *Baseline* case (2.104 s), CrossRoI reduces the latency by 25%. CrossRoI achieves less latency compared to the other alternatives because either less network overhead turns out to be less network delay or RoI-YOLO design boosts server inference speed. In this evaluation, we set the video streaming segment length as 1s. We notice that segment length is a critical parameter to system end-to-end response delay. We will provide more detailed discussion shortly in §5.3.

---
[5]We reduce the video resolutions to *540* p for server inference due to the lack of strong GPUs.

### 5.3 Sensitivity to Parameters

We investigate how three hyperparameters influence the performance of CrossRoI as follows:

(1) **SVM Model Non-Linearity**. We fine tune the $\gamma$ parameter to manipulate the non-linearity of the SVM filter model. A small $\gamma$ associates to a low non-linearity SVM kernel which usually can not fit training data perfectly and generates more outliers. A large $\gamma$ corresponds to a kernel model of high non-linearity which usually fits all the training data and cannot find outliers from the training samples.
(2) **RANSAC Threshold Distance**. In the regression filter, we manipulate the `residual-threshold` parameter of RANSAC, which determines the threshold distance for a sample to be regarded as an outlier. Specifically, we set `residual-threshold` $= \theta * mad$, where *mad* is the median absolute deviation of the training data and the default residual-threshold value of RANSAC algorithm. We fine tune different $\theta$ in the following evaluations instead.
(3) **Segment Length**. Segment length is the smallest temporal unit when cameras stream live videos to the CrossRoI server. Cameras compress all frames captured in the last segment in one shot before send it to the server. Segment length has significant influence on the network overhead and end-to-end latency.

The evaluation results are shown in figure 9, 10 and 11.

#### 5.3.1 SVM Model Non-Linearity.
As shown in figure 9, the system accuracy, network overhead and end-to-end response latency increase as $\gamma$ increases. A very small $\gamma$ causes the SVM Filters to remove too much negative outliers, which usually includes true negative samples. Hence, accuracy gets hurt when SVM non-linearity is very low. On the other hand, a small $\gamma$ leads to a smaller RoI mask for each camera as it removes negative ReID results fiercely, and thus, performs most significantly in reducing network overhead and end-to-end delay. We choose $\gamma = 10^{-4}$ in our system to achieve best system accuracy.

#### 5.3.2 RANSAC Threshold Distance.
As shown in figure 9, the system accuracy, network overhead and end-to-end response latency decrease as $\theta$ increases. A very low `residual-threshold` causes more positive ReID samples being detected as outliers, which usually leads to larger RoI regions for the cameras, and hence, improves the system accuracy but hurts its efficiency. We use $\theta = 0.01$ in our system to achieve highest system accuracy.

#### 5.3.3 Segment Length.
We present the network-latency trade-offs in figure 11 by tuning segment length parameter. segment length is a very significant impact factor for end-to-end response latency due to the queuing mechanism for video compression and streaming. Comparing to frame-by-frame image sending, chunked-video-based streaming causes data being queued at cameras memory, network interfaces and the server, and hence, increases the end-to-end latency. However, longer segment size provides better chance for cameras to compress the videos and significantly reduces the network overhead. We choose 1s segment length in CrossRoI to achieve least end-to-end delay.

### 5.4 Comparison & Integration with Frame Filtering Systems

As mentioned in §2, significant frame filtering works have been presented to alleviate resource contention for video analytics. For example, Reducto [27], the SotA frame filtering system, optimizes the cost/accuracy trade-offs by discarding frames in each segment when streaming videos from camera to the server. Such systems usually perform well when the query accuracy requirement is not high, e.g. counting vehicle numbers roughly to understand current traffic condition.

As CrossRoI exploits spatial redundancy in closely located camera-fleets, we treat frame filtering as an extra layer of optimization to augment our system when the query accuracy requirement is not very high (i.e. ≤ 95%). Specifically, we integrate Reducto into our system to build CrossRoI-Reducto. Similar to CrossRoI, Reducto also operates in two phases. It profiles video clips in offline phase to learn video patterns and applies the learned-patterns as frame filters to discard frames in online phase. It is natural to merge the two systems and generate CrossRoI-Reducto, which also operates in an offline-online mode.

The system workflow of CrossRoI-Reducto is shown in figure 12. In the offline phase, the CrossRoI module profiles offline video clips to generate *RoI masks*. Reducto module profiles "masks-cropped" offline video clips to learn the video patterns and generates frame filters for each camera. In the online phase, the video frames first go through RoI masks to remove spatial repentant content and then go through the frame filter to eliminate temporal redundancy. The remaining data is compressed by the video compressor as described in ❺ and sent to server for CNN inference.

Reducto can adjust how fiercely to filter the frames based on a given accuracy target (e.g. 90%). We set different accuracy targets from 85% to 100% and compare the system performance between Reducto and CrossRoI-Reducto. The evaluation results are presented in table 4. As shown in table 4, we measure the the frame-filtering capabilities of the two systems by showing how many frames are removed, from total 6000 frames (5 cameras × 120 seconds × 10 fps), in the video analytics process. Reducto and CrossRoI-Reducto removes different number of frames under same accuracy targets because full video and cropped videos exhibit different patterns, which Reducto depends on to filter frames. When we set the accuracy target as 100%, the frame filtering mechanism fails to work. Reducto degenerates to be the *Baseline* scenario and CrossRoI-Reducto degenerates to the original CrossRoI. In all other scenarios, both Reducto and CrossRoI-Reducto achieve the corresponding accuracy targets, while CrossRoI-Reducto outperforms Reducto in all three system performance metrics significantly, i.e. network overhead reduction by 48.3%, server throughput boosting by 1.45× and end-to-end latency reduction by 25.8%.

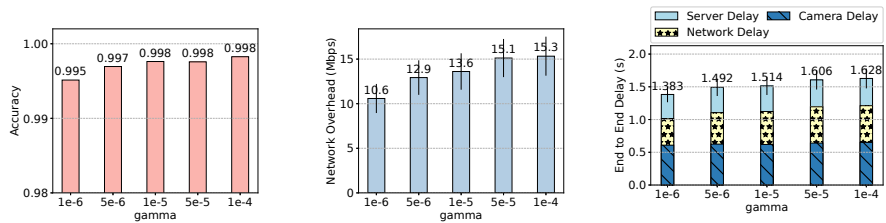

Figure 9: CrossRoI performance with different hyperparameter $\gamma$ (SVM model non-linearity).

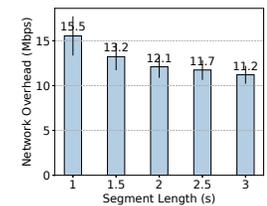

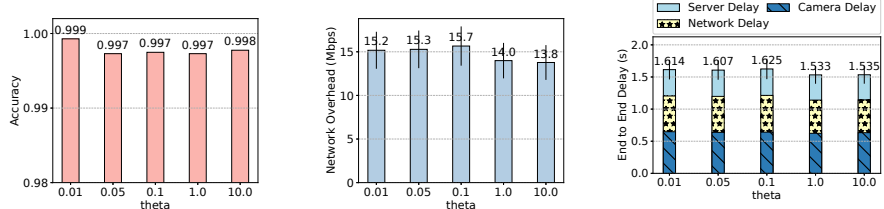

Figure 10: CrossRoI performance with different hyperparameter $\theta$ (RANSAC threshold distance).

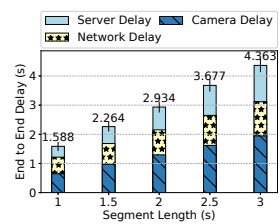

Figure 11: CrossRoI performance with different segment length.

| Accuracy Target | | Accuracy Achieved | Frames Reduced | Network Overhead (Mbps) | Server Throughput (Hz) | End-to-end Respond Latency (s) |
|---|---|---|---|---|---|---|
| Reducto | **1.00** | 1.000 | 0 | 26.48 | 52.07 | 2.104 |
|  | **0.95** | 0.971 | 979 | 23.85 | 62.32 | 1.884 |
|  | **0.90** | 0.947 | 2098 | 19.29 | 80.19 | 1.602 |
|  | **0.85** | 0.902 | 4116 | 10.16 | 166.01 | 1.063 |
| CrossRoI-Reducto | **1.00** | 0.999 | 0 | 15.73 (**-40.6%**) | 61.28 (**1.18 ×**) | 1.601 (**-23.9%**) |
|  | **0.95** | 0.962 | 1072 | 13.28 (**-44.3%**) | 74.17 (**1.19 ×**) | 1.406 (**-25.4%**) |
|  | **0.90** | 0.943 | 2389 | 10.48 (**-45.7%**) | 101.22 (**1.26 ×**) | 1.189 (**-25.8%**) |
|  | **0.85** | 0.893 | 4483 | 5.25 (**-48.3%**) | 240.95 (**1.45 ×**) | 0.821 (**-22.8%**) |

Table 4: Performance Comparison between Reducto and CrossRoI-Reducto. Bold number represents performance gains (server throughput) or resource reduction (network or latency overhead) of CrossRoI-Reducto compared to Reducto.

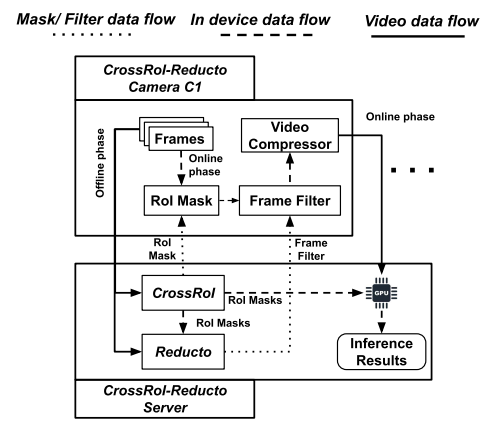

Figure 12: CrossRoI-Reducto system overview.

## 6 CONCLUSION

In this work, we present CrossRoI, a resource-efficient system that enables real time video analytics at scale via removing video content redundancy across a fleet of cameras. We develop a two-phase workflow in CrossRoI. In the offline phase, CrossRoI establishes cross-camera region associations to generate optimized RoI masks. In the online phase, CrossRoI applies the RoI masks to boost real time analytics performance. Experiments on real world traffic videos show that CrossRoI reduces network overhead by 42% ∼ 65% and reduces end-to-end response latency by 25% ∼ 34% when compared to baseline methods while maintaining 99.9% detection accuracy.

## ACKNOWLEDGMENT

This work was partially supported by the US Army Research Laboratory under cooperative agreement W911NF17-2-0196. The views and conclusions contained in this document are those of the authors and should not be interpreted as representing the official policies, either expressed or implied, of the Army Research Laboratory or the US government.

## REFERENCES

[1] [n. d.]. 45 Billion Cameras by 2022 Fuel Business Opportunities. https://www.ldv.co/insights/2017. Accessed: 2021-01-27.
[2] [n. d.]. Absolutely everywhere in Beijing is now covered by police video surveillance. https://qz.com/518874/. Accessed: 2021-01-27.
[3] [n. d.]. Arecont Vision MegaVideo UltraHD. https://sales.arecontvision.com/product/MegaVideo+UltraHD+Series/AV12ZMV-301. Accessed: 2021-01-27.
[4] [n. d.]. British transport police: CCTV. http://www.btp.police.uk/advice_and_information/safety_on_and_near_the_railway/cctv.aspx. Accessed: 2021-01-27.
[5] [n. d.]. Can 30000 Cameras Help Solve Chicago's Crime Problem? https://www.nytimes.com/2018/05/26/us/chicago-police-surveillance.html. Accessed: 2021-01-27.
[6] [n. d.]. FFmpeg. https://ffmpeg.org. Accessed: 2021-01-27.
[7] [n. d.]. Gurobi Solver. https://www.gurobi.com/. Accessed: 2021-01-27.
[8] [n. d.]. Logitech C930e BUSINESS WEBCAM. https://www.logitech.com/en-us/products/webcams/. Accessed: 2021-01-27.
[9] [n. d.]. Scikit Learn. https://scikit-learn.org/. Accessed: 2021-01-27.
[10] [n. d.]. Tensorflow. https://www.tensorflow.org/. Accessed: 2021-01-27.
[11] Christopher Canel, Thomas Kim, Giulio Zhou, Conglong Li, Hyeontaek Lim, David G Andersen, Michael Kaminsky, and Subramanya R Dulloor. 2019. Scaling video analytics on constrained edge nodes. *arXiv preprint arXiv:1905.13536* (2019).
[12] Bo Chen, Zhisheng Yan, Haiming Jin, and Klara Nahrstedt. 2019. Event-driven stitching for tile-based live 360 video streaming. In *Proceedings of the 10th ACM Multimedia Systems Conference (MMsys)*. 1–12.
[13] Tiffany Yu-Han Chen, Lenin Ravindranath, Shuo Deng, Paramvir Bahl, and Hari Balakrishnan. 2015. Glimpse: Continuous, real-time object recognition on mobile devices. In *Proceedings of the 13th ACM Conference on Embedded Networked Sensor Systems (Sensys)*. 155–168.

[14] Sandeep P Chinchali, Eyal Cidon, Evgenya Pergament, Tianshu Chu, and Sachin Katti. 2018. Neural networks meet physical networks: Distributed inference between edge devices and the cloud. In *Proceedings of the 17th ACM Workshop on Hot Topics in Networks (HotNets)*. 50–56.

[15] Kuntai Du, Ahsan Pervaiz, Xin Yuan, Aakanksha Chowdhery, Qizheng Zhang, Henry Hoffmann, and Junchen Jiang. 2020. Server-Driven Video Streaming for Deep Learning Inference. In *Proceedings of the Annual conference of the ACM Special Interest Group on Data Communication (SIGCOMM)*. 557–570.

[16] Zhou Fang, Dezhi Hong, and Rajesh K Gupta. 2019. Serving deep neural networks at the cloud edge for vision applications on mobile platforms. In *Proceedings of the 10th ACM Multimedia Systems Conference (MMsys)*. 36–47.

[17] Xianglong Feng, Viswanathan Swaminathan, and Sheng Wei. 2019. Viewport prediction for live 360-degree mobile video streaming using user-content hybrid motion tracking. *Proceedings of the ACM on Interactive, Mobile, Wearable and Ubiquitous Technologies (IMWUT)* 3, 2 (2019), 1–22.

[18] Martin A Fischler and Robert C Bolles. 1981. Random sample consensus: a paradigm for model fitting with applications to image analysis and automated cartography. *Commun. ACM* 24, 6 (1981), 381–395.

[19] Yu Guan, Chengyuan Zheng, Xinggong Zhang, Zongming Guo, and Junchen Jiang. 2019. Pano: Optimizing 360 video streaming with a better understanding of quality perception. In *Proceedings of the ACM Special Interest Group on Data Communication (SIGCOMM)*. 394–407.

[20] Anhong Guo, Anuraag Jain, Shomiron Ghose, Gierad Laput, Chris Harrison, and Jeffrey P Bigham. 2018. Crowd-ai camera sensing in the real world. *Proceedings of the ACM on Interactive, Mobile, Wearable and Ubiquitous Technologies (IMWUT)* 2, 3 (2018), 1–20.

[21] Zhiqun He, Yu Lei, Shuai Bai, and Wei Wu. 2019. Multi-Camera Vehicle Tracking with Powerful Visual Features and Spatial-Temporal Cue.. In *CVPR Workshops*. 203–212.

[22] Marti A. Hearst, Susan T Dumais, Edgar Osuna, John Platt, and Bernhard Scholkopf. 1998. Support vector machines. *IEEE Intelligent Systems and their applications* 13, 4 (1998), 18–28.

[23] Kevin Hsieh, Ganesh Ananthanarayanan, Peter Bodik, Shivaram Venkataraman, Paramvir Bahl, Matthai Philipose, Phillip B Gibbons, and Onur Mutlu. 2018. Focus: Querying large video datasets with low latency and low cost. In *13th USENIX Symposium on Operating Systems Design and Implementation (OSDI)*. 269–286.

[24] Samvit Jain, Ganesh Ananthanarayanan, Junchen Jiang, Yuanchao Shu, and Joseph Gonzalez. 2019. Scaling video analytics systems to large camera deployments. In *Proceedings of the 20th International Workshop on Mobile Computing Systems and Applications (HotMobile)*. 9–14.

[25] Samvit Jain, Xun Zhang, Yuhao Zhou, Ganesh Ananthanarayanan, Junchen Jiang, Yuanchao Shu, Victor Bahl, and Joseph Gonzalez. 2020. Spatula: Efficient cross-camera video analytics on large camera networks. In *ACM/IEEE Symposium on Edge Computing (SEC)*.

[26] Peilun Li, Guozhen Li, Zhangxi Yan, Youzeng Li, Meiqi Lu, Pengfei Xu, Yang Gu, Bing Bai, Yifei Zhang, and DiDi Chuxing. 2019. Spatio-temporal Consistency and Hierarchical Matching for Multi-Target Multi-Camera Vehicle Tracking.. In *CVPR Workshops*. 222–230.

[27] Yuanqi Li, Arthi Padmanabhan, Pengzhan Zhao, Yufei Wang, Guoqing Harry Xu, and Ravi Netravali. 2020. Reducto: On-Camera Filtering for Resource-Efficient Real-Time Video Analytics. In *Proceedings of the Annual conference of the ACM Special Interest Group on Data Communication (SIGCOMM)*. 359–376.

[28] Bingyan Liu, Yuanchun Li, Yunxin Liu, Yao Guo, and Xiangqun Chen. 2020. PMC: A Privacy-preserving Deep Learning Model Customization Framework for Edge Computing. *Proceedings of the ACM on Interactive, Mobile, Wearable and Ubiquitous Technologies (IMWUT)* 4, 4 (2020), 1–25.

[29] Wei Liu, Dragomir Anguelov, Dumitru Erhan, Christian Szegedy, Scott Reed, Cheng-Yang Fu, and Alexander C Berg. 2016. SSD: Single shot multibox detector. In *European conference on computer vision (ECCV)*. Springer, 21–37.

[30] Xiaochen Liu, Pradipta Ghosh, Oytun Ulutan, BS Manjunath, Kevin Chan, and Ramesh Govindan. 2019. Caesar: cross-camera complex activity recognition. In *Proceedings of the 17th Conference on Embedded Networked Sensor Systems (Sensys)*. 232–244.

[31] Xinchen Liu, Wu Liu, Huadong Ma, and Huiyuan Fu. 2016. Large-scale vehicle re-identification in urban surveillance videos. In *2016 IEEE International Conference on Multimedia and Expo (ICME)*. IEEE, 1–6.

[32] Milind Naphade, Zheng Tang, Ming-Ching Chang, David C Anastasiu, Anuj Sharma, Rama Chellappa, Shuo Wang, Pranamesh Chakraborty, Tingting Huang, Jenq-Neng Hwang, et al. 2019. The 2019 AI City Challenge.. In *CVPR Workshops*. 452–460.

[33] Xukan Ran, Haolianz Chen, Xiaodan Zhu, Zhenming Liu, and Jiasi Chen. 2018. Deepdecision: A mobile deep learning framework for edge video analytics. In *2018-IEEE Conference on Computer Communications (INFOCOM)*. IEEE, 1421–1429.

[34] Joseph Redmon, Santosh Divvala, Ross Girshick, and Ali Farhadi. 2016. You only look once: Unified, real-time object detection. In *Proceedings of the IEEE conference on computer vision and pattern recognition (CVPR)*. 779–788.

[35] Joseph Redmon and Ali Farhadi. 2018. Yolov3: An incremental improvement. *arXiv preprint arXiv:1804.02767* (2018).

[36] Mengye Ren, Andrei Pokrovsky, Bin Yang, and Raquel Urtasun. 2018. Sbnet: Sparse blocks network for fast inference. In *Proceedings of the IEEE Conference on Computer Vision and Pattern Recognition (CVPR)*. 8711–8720.

[37] Shaoqing Ren, Kaiming He, Ross Girshick, and Jian Sun. 2015. Faster r-cnn: Towards real-time object detection with region proposal networks. *arXiv preprint arXiv:1506.01497* (2015).

[38] Ergys Ristani, Francesco Solera, Roger Zou, Rita Cucchiara, and Carlo Tomasi. 2016. Performance measures and a data set for multi-target, multi-camera tracking. In *European conference on computer vision (ECCV)*. Springer, 17–35.

[39] Ergys Ristani and Carlo Tomasi. 2018. Features for multi-target multi-camera tracking and re-identification. In *Proceedings of the IEEE conference on computer vision and pattern recognition (CVPR)*. 6036–6046.

[40] Gary J Sullivan, Jens-Rainer Ohm, Woo-Jin Han, and Thomas Wiegand. 2012. Overview of the high efficiency video coding (HEVC) standard. *IEEE Transactions on circuits and systems for video technology* 22, 12 (2012), 1649–1668.

[41] Junjue Wang, Ziqiang Feng, Zhuo Chen, Shilpa George, Mihir Bala, Padmanabhan Pillai, Shao-Wen Yang, and Mahadev Satyanarayanan. 2018. Bandwidth-efficient live video analytics for drones via edge computing. In *2018 IEEE/ACM Symposium on Edge Computing (SEC)*. IEEE, 159–173.

[42] Qianru Wang, Junbo Zhang, Bin Guo, Zexia Hao, Yifang Zhou, Junkai Sun, Zhiwen Yu, and Yu Zheng. 2019. CityGuard: citywide fire risk forecasting using a machine learning approach. *Proceedings of the ACM on Interactive, Mobile, Wearable and Ubiquitous Technologies (IMWUT)* 3, 4 (2019), 1–21.

[43] Thomas Wiegand, Gary J Sullivan, Gisle Bjontegaard, and Ajay Luthra. 2003. Overview of the H. 264/AVC video coding standard. *IEEE Transactions on circuits and systems for video technology* 13, 7 (2003), 560–576.

[44] Mengbai Xiao, Chao Zhou, Yao Liu, and Songqing Chen. 2017. Optile: Toward optimal tiling in 360-degree video streaming. In *Proceedings of the 25th ACM international conference on Multimedia (ACM MM)*. 708–716.

[45] Shanhe Yi, Zijiang Hao, Qingyang Zhang, Quan Zhang, Weisong Shi, and Qun Li. 2017. Lavea: Latency-aware video analytics on edge computing platform. In *2017 ACM/IEEE Symposium on Edge Computing (SEC)*. 1–13.

[46] Haoyu Zhang, Ganesh Ananthanarayanan, Peter Bodik, Matthai Philipose, Paramvir Bahl, and Michael J Freedman. 2017. Live video analytics at scale with approximation and delay-tolerance. In *14th USENIX Symposium on Networked Systems Design and Implementation (NSDI)*. 377–392.

[47] Shigeng Zhang, Yinggang Li, Xuan Liu, Song Guo, Weiping Wang, Jianxin Wang, Bo Ding, and Di Wu. 2020. Towards Real-time Cooperative Deep Inference over the Cloud and Edge End Devices. *Proceedings of the ACM on Interactive, Mobile, Wearable and Ubiquitous Technologies (IMWUT)* 4, 2 (2020), 1–24.

[48] Liang Zheng, Yi Yang, and Alexander G Hauptmann. 2016. Person re-identification: Past, present and future. *arXiv preprint arXiv:1610.02984* (2016).